\pdfoutput=1

\documentclass[11pt,twoside,a4paper,cmspaper,final,collab]{cms-tdr}
\def\svnVersion{297817}\def\cmsCernNo{2013-037}\def\cmsCernDate{\today}\def\cmsMessage{Published in Physics Letters B as \href{http://dx.doi.org/10.1016/j.physletb.2015.07.010}{\doi{10.1016/j.physletb.2015.07.010.}}}

\begin{document}\cmsNoteHeader{HIG-14-011}

\hyphenation{had-ron-i-za-tion}
\hyphenation{cal-or-i-me-ter}
\hyphenation{de-vices}
\RCS$Revision: 297804 $
\RCS$HeadURL: svn+ssh://svn.cern.ch/reps/tdr2/papers/HIG-14-011/trunk/HIG-14-011.tex $
\RCS$Id: HIG-14-011.tex 297804 2015-07-23 12:59:03Z zucchett $
\newlength\cmsFigWidth
\ifthenelse{\boolean{cms@external}}{\setlength\cmsFigWidth{0.95\columnwidth}}{\setlength\cmsFigWidth{0.45\textwidth}}
\newlength\cmsFigWidthOneSix
\ifthenelse{\boolean{cms@external}}{\setlength\cmsFigWidthOneSix{0.95\columnwidth}}{\setlength\cmsFigWidthOneSix{0.75\textwidth}}
\ifthenelse{\boolean{cms@external}}{\providecommand{\cmsLeft}{top}}{\providecommand{\cmsLeft}{left}}%
\ifthenelse{\boolean{cms@external}}{\providecommand{\cmsRight}{bottom}}{\providecommand{\cmsRight}{right}}

\newcommand{\A}{\ensuremath{\cmsSymbolFace{A}}\xspace}
\renewcommand{\PW}{\ensuremath{\cmsSymbolFace{W}}\xspace}
\newcommand{\AtoZh}{\ensuremath{\A\to\Z\Ph}\xspace}
\newcommand{\htobb}{\ensuremath{\Ph\to\bbbar}\xspace}
\newcommand{\Ztoll}{\ensuremath{\Z\to\ell\ell}\xspace}
\newcommand{\Atollbb}{\ensuremath{\A\to\Z\Ph\to{\ell\ell}\bbbar}\xspace}
\renewcommand{\PHpm}{\ensuremath{\PH^\pm}\xspace}
\newcommand{\mA}{\ensuremath{m_{\A}}\xspace}
\newcommand{\mh}{\ensuremath{m_{\Ph}}\xspace}
\newcommand{\mH}{\ensuremath{m_{\PH}}\xspace}
\newcommand{\mHpm}{\ensuremath{m_{\PHpm}}\xspace}
\newcommand{\mt}{\ensuremath{m_{\PQt}}\xspace}
\newcommand{\ttbarV}{\ensuremath{\ttbar V\xspace}}
\newcommand{\mbb}{\ensuremath{m_{\PQb\PQb}}\xspace}
\newcommand{\mjj}{\ensuremath{m_{jj}}\xspace}
\newcommand{\mll}{\ensuremath{m_{\ell\ell}}\xspace}
\newcommand{\mllbb}{\ensuremath{m_{\ell\ell\PQb\PQb}}\xspace}
\newcommand{\cosba}{\ensuremath{\cos(\beta-\alpha)}\xspace}
\newcommand{\ST}{\ensuremath{S_{\mathrm{T}}}\xspace}
\newcommand{\B}{\ensuremath{\mathcal{B}}}
\newcommand{\fb}{\ensuremath{\,\text{fb}}\xspace}
\newcommand{\Zb}{\ensuremath{\Z+\PQb}\xspace}
\newcommand{\Zbbbar}{\ensuremath{\Z+\bbbar}\xspace}

\cmsNoteHeader{HIG-14-011}
\title{Search for a pseudoscalar boson decaying into a Z boson and the 125\GeV Higgs boson in $\ell^+ \ell^- \bbbar$ final states}

\date{\today}

\abstract{
Results are reported on a search for decays of a pseudoscalar A boson into a Z boson and a light scalar h boson, where the Z boson decays into a pair of oppositely-charged electrons or muons, and the h boson decays into $\bbbar$. The search is based on data from proton-proton collisions at a center-of-mass energy $\sqrt{s}=8\TeV$ collected with the CMS detector, corresponding to an integrated luminosity of 19.7\fbinv. The h boson is assumed to be the standard model-like Higgs boson with a mass of 125\GeV. With no evidence for signal, upper limits are obtained on the product of the production cross section and the branching fraction of the A boson in the Zh channel. Results are also interpreted in the context of two Higgs doublet models.
}
\hypersetup{%
pdfauthor={CMS Collaboration},%
pdftitle={Search for a pseudoscalar boson decaying into a Z boson and the 125 GeV Higgs boson in llbb final states},%
pdfsubject={CMS},%
pdfkeywords={CMS, Higgs, Pseudoscalar, Lepton, b-quark, 2HDM}
}

\maketitle

\section{Introduction}

The discovery of a scalar boson at the CERN LHC~\cite{bib:Aad20121,bib:Chatrchyan201230,Chatrchyan:2013lba} with properties in agreement with those predicted by the standard model (SM) raises the question of whether the Higgs sector consists of only one physical state, as expected in the SM, or whether additional bosons are also involved.

An extension of the SM Higgs sector is provided in two Higgs doublet models (2HDM)~\cite{bib:Branco20121}, which introduce a second scalar doublet in addition to the one from the SM.
Different formulations of 2HDM predict different couplings of the two doublets to quarks and to leptons.
In Type-I 2HDM, all fermions couple to only one Higgs doublet, while in Type-II, up- and down-type quarks couple to different doublets. One example of a
Type-II 2HDM is the minimal supersymmetric standard model~\cite{Martin:1997ns}, despite that supersymmetry is not explicitly required in 2HDM.
The second Higgs doublet entails the presence of five physical states: two neutral, CP-even states, representing a light \Ph and a heavy \PH boson; a neutral, CP-odd \A boson; and two charged scalar \PHpm bosons. The lightest scalar \Ph is assumed to be the boson observed at the LHC at a mass of
125\GeV~\cite{Khachatryan:2014jba,Aad:2014aba,Aad:2013xqa,Aad:2015zhl}.
If the masses of the heavier bosons are at or below the \TeV scale, they can be accessible at the LHC.
Searches for this extended sector can be performed either by measuring the values
of the couplings of the discovered \Ph boson to other SM particles~\cite{bib:Craig:2012vn,bib:Craig:2013hca,bib:Ferreira:2013qua},
or via direct searches in final states disfavored by the SM~\cite{bib:Khachatryan:2014jya}.
A way to probe this kind of new physics is therefore to search for bosons that decay into final states that contain an SM-like Higgs boson.

This paper describes a search for a heavy pseudoscalar \A boson that decays into a \Z and an \Ph boson, both on-shell, with the \Z boson decaying into a pair of $\ell^+\ell^-$ leptons ($\ell$ being $\Pe$ or $\mu$), and the \Ph boson into \bbbar.
In most 2HDM formulations~\cite{bib:Branco20121}, the \A boson is produced predominantly through gluon-gluon fusion and decays to on-shell \Z and \Ph bosons, provided that the mass of the \A boson satisfies $\mA \gtrsim \mh+m_\Z \approx 216\GeV$.
This channel is expected to be viable for $\mA$ smaller than twice the top quark mass (\mt), where the decay \AtoZh is generally dominant, but, depending on model parameters, it can also be sensitive at larger values of \mA. The \htobb decay has a large branching fraction for most of the parameter space in 2HDM~\cite{bib:Craig:2013hca}.
A similar analysis has been recently published by ATLAS~\cite{Aad:2015wra}.

The analysis strategy is to reconstruct the \Z, \Ph, and \A boson candidates from the visible decay products in the event. The signal would manifest itself as a peak in the four-body invariant mass (\mllbb) spectrum over an expected SM continuum.
Irreducible backgrounds correspond to \Z boson production with two accompanying b quark jets, and \ttbar events in the dileptonic final state.
These backgrounds are evaluated and normalized directly using appropriate control regions in data.
The \Ph boson produced in association with a \Z boson provides a contribution to the background, but it differs from signal because the \mllbb mass does not contain a resonant peak.
Signal sensitivity is improved by exploiting the known value of the \Ph boson mass, using it to rescale the jet momenta to match the value expected for the dijet invariant mass. In addition, optimal signal efficiency and background rejection is achieved using a multivariate discriminator. Results are extracted through a two-dimensional (2D) fit to \mllbb and the discriminator output; upper limits are presented on the product of the total cross section and the $\A\to\Z\Ph$, $\Z\to\ell\ell$, and $\Ph\to\bbbar$ branching fractions for a pseudoscalar boson, and interpreted within the 2HDM.

\section{CMS detector}\label{sec:detector}

A detailed description of the CMS detector, together with a definition of the coordinate system and kinematic variables, can be found in Ref.~\cite{Chatrchyan:2008zzk}.
The central feature of the CMS apparatus is a superconducting solenoid of 6\unit{m} internal diameter, providing a magnetic field of 3.8\unit{T}. The field volume contains a silicon pixel and strip tracker, a homogeneous electromagnetic calorimeter (ECAL), and a sampling hadron calorimeter (HCAL), each composed of a barrel and two endcap sections. Muons are measured in gas-ionization detectors embedded in the steel flux-return yoke outside the solenoid.

The silicon tracker measures charged particles within the pseudorapidity range of $\abs{\eta}< 2.5$.
For non-isolated particles with transverse momenta of $1 < \pt < 10\GeV$ and $\abs{\eta} < 1.4$, the track resolutions are typically 1.5\% in \pt, and between 25--90 and 45--150\mum, respectively, in transverse and longitudinal impact parameters relative to the production vertex~\cite{TRK-11-001}.
The ECAL consists of lead tungstate crystals that provide a coverage up to $\abs{\eta} < 3.0$.
The mass resolution for $\Z \to \Pep \Pem$ decays when both electrons are in the ECAL barrel is 1.6\%, and is 2.6\% when both electrons are in the endcaps.
The HCAL has alternating layers of brass as absorber and plastic scintillators, and
covers the range of $\abs{\eta} < 3.0$, which is extended to $\abs{\eta} \lesssim 5.2$ through forward calorimetry.
Muons are measured in the range of $\abs{\eta}< 2.4$, with detection planes made using three technologies: drift tubes, cathode-strip chambers, and resistive-plate chambers. Matching muons to tracks measured in the silicon tracker provides a \pt resolution of 1.3--2.0\% for muons with $20 <\pt < 100\GeV$ in the barrel and better than 6\% in the endcaps.

\section{Data and simulation}\label{sec:data}

Data used for this analysis were collected using double-muon and double-electron triggers, with \pt thresholds set to 17 and 8\GeV for the highest and next highest \pt lepton, respectively, and an isolation requirement used in the electron trigger to maintain an acceptable rate.
The analyzed events correspond to an integrated luminosity of $19.7 \pm 0.5\fbinv$ of pp collisions at $\sqrt{s}=8\TeV$~\cite{lumi}.

Signal samples and SM background processes \Z{}+jets, \PW{}+jets, and \ttbar{}+jets or a vector boson (\ttbarV) are simulated using the \MADGRAPH 5.1~\cite{ref:MADGRAPH} Monte Carlo (MC) generator; multijets and dibosons ($VV^{'}$, with $V$, $V^{'}$ being \PW or \Z) are generated using LO \PYTHIA 6.4~\cite{Sjostrand:2006za} and the CTEQ6L~\cite{CTEQ6} parton distribution functions (PDF).
Single top quark production, and SM Higgs boson production in association with an electroweak vector boson ($V$\Ph) are generated using the next-to-leading-order (NLO) \POWHEG 1.0~\cite{powheg1,powheg2,powheg3} MC generator and the MSTW2008NLO PDF~\cite{MSTW2008}.
Parton showering and hadronization are performed with \PYTHIA using the Z2* tune~\cite{Field:2010bc}.
The generated MC events, including additional pp interactions (pileup) occurring in the bunch crossing containing the high-\pt scatter, are processed through a full detector simulation based on \GEANTfour~\cite{ref:GEANT4} and reconstructed with the same algorithms as used for data.

The pseudoscalar \A boson is assumed to be produced via the gluon-gluon fusion process and to have a narrow width.
The validity of the narrow-width approximation is discussed in Section~\ref{sec:results}.
The branching fraction $\B(\AtoZh)$ is set to 100\%, and similarly only the \Z boson decays into electrons or muons, and the \Ph boson decays
into a pair of \bbbar quarks, are considered.
Other decay modes have negligible efficiency for passing the selections.
The mass of the light Higgs boson is set to \mbox{$m_{\Ph}=125\GeV$}, while the search for the \A boson is performed in the mass range from 225 to 600\GeV, above which the efficiency to reconstruct two separate jets from \Ph decay becomes too small because of its increased momentum.

\section{Event reconstruction}\label{sec:reco}

A global reconstruction of the event is achieved using a particle-flow (PF) technique, which reconstructs and identifies individual particles emerging from each collision using information from all CMS subdetectors~\cite{bib:CMS-PAS-PFT-09-001,bib:CMS-PAS-PFT-10-001}.

Electron candidates are reconstructed for $\abs{\eta}<2.5$ by matching
energy depositions in the ECAL with reconstructed tracks~\cite{Khachatryan:2015hwa}.
The identification relies on a multivariate technique that combines observables
sensitive to the amount of bremsstrahlung along the electron trajectory,
the compatibility of the measured position, direction and momentum of the reconstructed track in the inner tracker, and the energy deposition reconstructed in the ECAL cluster.
Additional requirements are imposed to remove electrons produced by photon conversions in the detector material.

Muons are reconstructed within $\abs{\eta}<2.4$, combining
information from both the silicon tracker and the outer muon spectrometer~\cite{Chatrchyan:2012xi}, requiring small energy depositions in the calorimeters~\cite{CMS-PAS-PFT-10-003}.
Muon candidates have to fulfill restrictive selection criteria based on the quality and the impact parameter of the track, as well as on the number of hits observed in the tracker and muon systems.

An isolation variable is defined for each lepton through the scalar sum over the \pt of all PF candidates, excluding the lepton,
within a cone of $\Delta R= \sqrt{\smash[b]{(\Delta \eta)^2 + (\Delta \varphi)^2}}<0.4$ around the lepton direction, where $\varphi$ is the azimuthal angle in
radians, subtracting the pileup contribution~\cite{Cacciari2008119}, and then dividing by the lepton \pt. The lepton is rejected if the
isolation exceeds 0.15 for electrons and 0.12 for muons.

Jets are formed from all particles, charged and neutral, reconstructed through the PF
algorithm, and clustered with the anti-\kt
algorithm~\cite{Cacciari:2008gp,Cacciari:2011ma} with a distance parameter of 0.5.
Jet energy corrections are determined from dijet and $\Z/\gamma$+jet events, and applied to both data and simulation~\cite{CMS-PAS-JME-10-010}.
The imbalance in transverse momentum is calculated as the negative of the vectorial sum of transverse momenta of all the PF candidates, and its magnitude is denoted as \MET~\cite{1748-0221-10-02-P02006}.

The combined secondary vertex (CSV) b tagging
algorithm~\cite{CMS-PAS-BTV-13-001} is used to identify jets that
originate from b quarks. This algorithm combines information from track
impact parameters relative to the primary and secondary (displaced) vertices
into a likelihood discriminant.
Two standard working points are set on the discriminant, corresponding to restrictive (tight) and less-restrictive (loose)
thresholds that provide on average 50\% and 80\% b tagging efficiencies,
with respective misidentification probabilities of about 0.1\% and 10\% for light-flavor jets, and about 5\% and 25\% for c jets.
The CSV distribution is corrected in simulation to take into account a difference at the percent level in algorithm performance for data and simulation.

\section{Event selection}\label{sec:selection}

Events are required to have at least two electrons or two muons within the geometrical
acceptance regions, and to satisfy the reconstruction, identification,
and isolation requirements. The \pt threshold is set to
20\GeV for the lepton with highest \pt, and to 10\GeV for the lepton with next-highest \pt. The
\Z boson candidate is formed from the two highest-\pt, opposite-charge, same-flavor leptons, and must
have an invariant mass larger than 50\GeV. In addition, at least two jets are
required with $\pt>20\GeV$, within $\abs{\eta}<2.4$, and with angular
separation relative to each lepton of $\Delta R>0.5$. The \Ph boson
candidate is formed from the two jets that have the highest values of the b tagging
discriminant.
Additional jets in the event are ignored.

Signal events are expected to have at least one jet b-tagged with the tight and one with the loose CSV working points;
after b tagging application, the contribution from pileup jets becomes negligible.
The invariant masses of the two leptons and of the two jets are required to be compatible with those of the \Z boson ($75<\mll<105\GeV$) and \Ph boson ($90<\mbb<140\GeV$), and the value of the \MET in the event has to be compatible with zero. After these selections, the signal efficiencies range from $10.0\%$ for $\mA=250 \GeV$ to $20.7\%$ for $\mA=600 \GeV$.

Dedicated control regions are defined to check both the normalizations and distributions of the most important backgrounds by inverting the selections used to enhance signal.
Drell--Yan backgrounds (from $\qqbar \to \Z/\gamma^\star \to \ell^+ \ell^-$ production) are considered separately as a function of the number of b jets, distinguishing \Z{}+jets (no b jets), \Zb (1 b jet) and \Zbbbar (2 b jets). The corresponding control regions are selected by requiring $80<\mll<100\GeV$, $\MET<40\GeV$, vetoing of dijet masses close to the Higgs boson mass of $90<\mjj<140\GeV$, and applying different b-tagging selections.
In the \Z{}+jets control region, no cutoffs are applied on b tagging discriminators.
The \Zb control region contains events with one b jet fulfilling the tight CSV working point, and no other jets passing the loose threshold.
Events that enter the \Zbbbar control region should have at least two b-tagged jets, with one passing the tight and the other the loose working point.
The \ttbar control region is defined by inverting the \mll and \MET selections, dropping the
requirement of the dijet mass, and requiring at least one tight and one loose b-tagged jet.

The scale factors that are used to correct the normalization of Drell--Yan and \ttbar backgrounds, reported in Table~\ref{tab:CR}, are
obtained from a simultaneous likelihood fit to data and simulation in the
four control regions and are applied in the following steps of the analysis.
Multijet contamination in control and signal regions is evaluated with data by inverting lepton isolation criteria, and its contribution is found to be negligible.
The yield of the $V$h, \ttbarV, and single top production through the $s$ channel are calculated at NLO~\cite{Alioli:2009je,Dittmaier:2012vm}, while the other single top channels and dibosons are normalized to the measured cross sections~\cite{CMS-PAPERS-TOP-12-036,PhysRevLett.112.231802,tagkey2013190}.
A mismodeling in simulation, observed in the control regions relative to data, is corrected by reweighting with a linear function the event centrality distribution (defined as the ratio of the sums in scalar \pt and the energy of the two leptons and two jets in the rest frame of the four objects) in all Monte Carlo events, improving the overall agreement between simulation and data.

\begin{table}[!htb]
    \topcaption{Scale factors for the four main backgrounds obtained from a fit to the control regions. Reported uncertainties are statistical in nature.}\label{tab:CR}
    \centering
    \ifthenelse{\boolean{cms@external}}{
        \begin{tabular}{lr}
            \hline
            Background & Scale factor\\
            \hline
            $\Z+\text{jets}$ & $1.069 \pm 0.002$ \\
            \Zb & $0.945 \pm 0.012$ \\
            \Zbbbar & $1.008 \pm 0.020$ \\
            \ttbar & $0.984 \pm 0.010$\\
            \hline
        \end{tabular}
    }
    {
        \begin{tabular}{lcccc}
            \hline
            Background & $\Z+\text{jets}$ & \Zb & \Zbbbar & \ttbar \\
            \hline
            Scale factor & $1.069 \pm 0.002$ & $0.945 \pm 0.012$ & $1.008 \pm 0.020$ & $0.984 \pm 0.010$\\
            \hline
        \end{tabular}
    }
\end{table}

\begin{figure}[!htb]\centering
  \includegraphics[width=\cmsFigWidthOneSix]{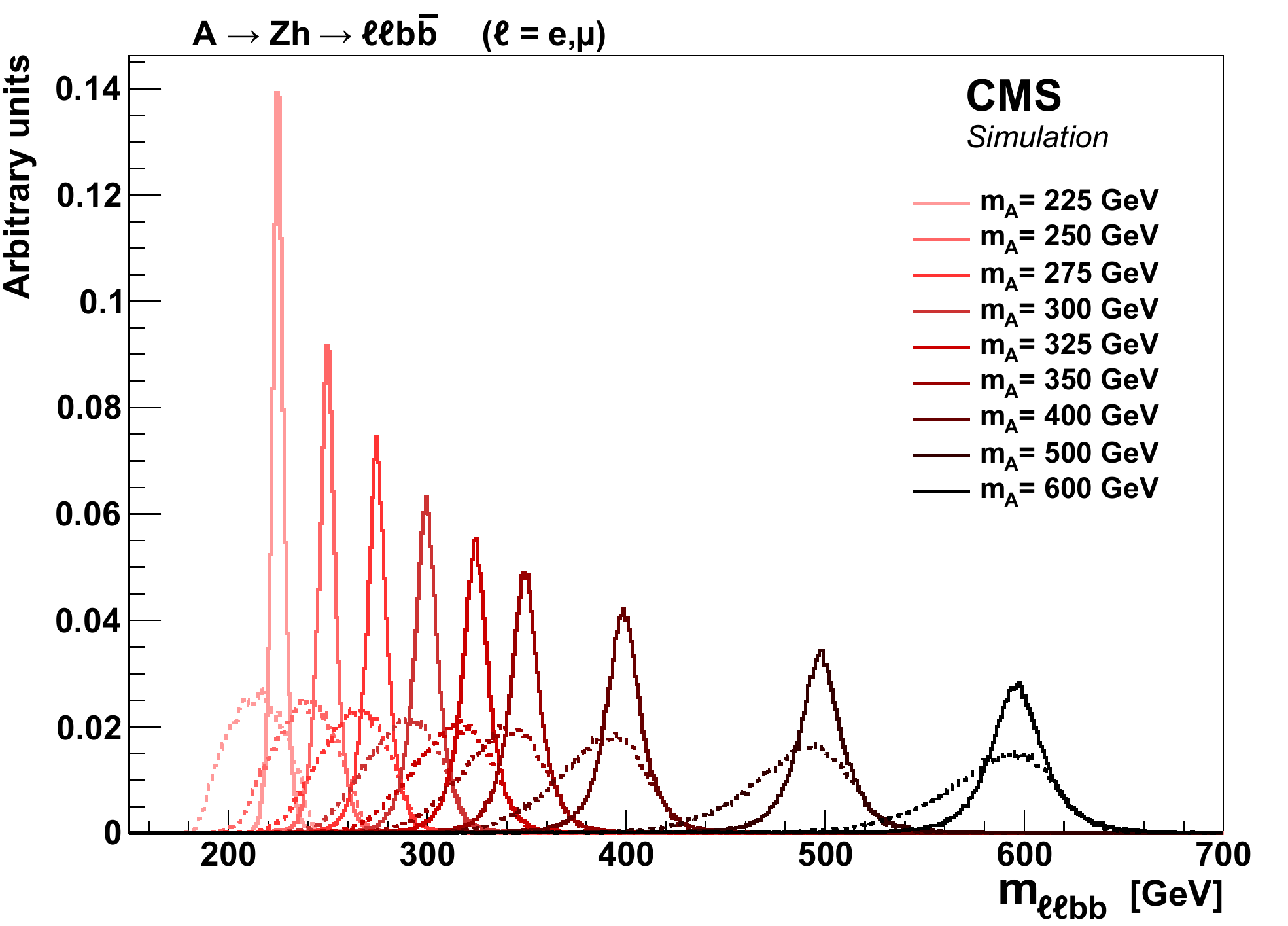}
  \caption{Simulated distributions for \mllbb before (dotted lines) and after the kinematic fits (solid lines). Histograms are normalized to unit area.}
  \label{fig:KinFit}
\end{figure}

An important feature of the signal is that the two b jets originate
from the decay of the \Ph boson, whose mass is known with better
precision than that provided by the \bbbar invariant mass resolution~\cite{PhysRevD.89.012003}.
The measured jet \pt, $\eta$, and $\varphi$ values are therefore varied according to their
resolution, in a kinematic fit based on Lagrange multipliers, to constrain the dijet
invariant mass to $\mh=125\GeV$. The fit $\chi^2$ is used in
subsequent steps of the analysis as a discriminant in place of \mbb.
The kinematic fit improves the relative four-body invariant mass resolution
from 6.3\% to 1.2\% and 4.0\% to 1.9\%, respectively, for the smallest and largest values of \mA,
centering the peaks around
their nominal values, as shown in Fig.~\ref{fig:KinFit}.
The effect of the kinematic fit is larger at low \mA, where the constraint on the Higgs invariant mass has the largest contribution.
Although both the background and signal \mllbb
distributions are modified by the kinematic fit, the signal significance in a
mass window close to the investigated \A boson mass increases by a factor of two at the lowest mass
and by 34\% at the highest mass. The resulting jet three-momenta are used
to redefine all the kinematic variables in the event.

\begin{figure*}[!hbtp]\centering
  \includegraphics[width=0.95\cmsFigWidth]{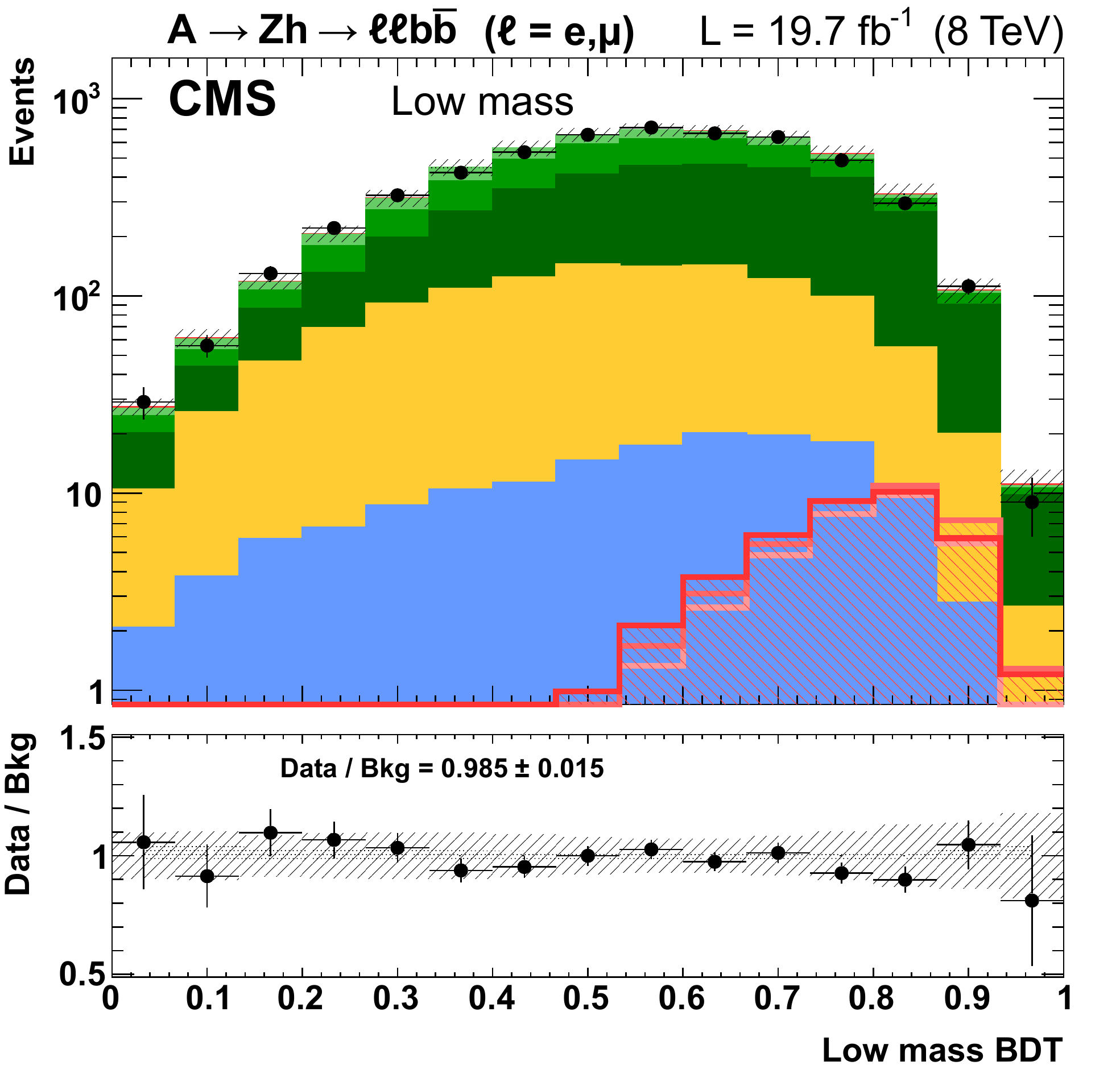}
  \includegraphics[width=0.95\cmsFigWidth]{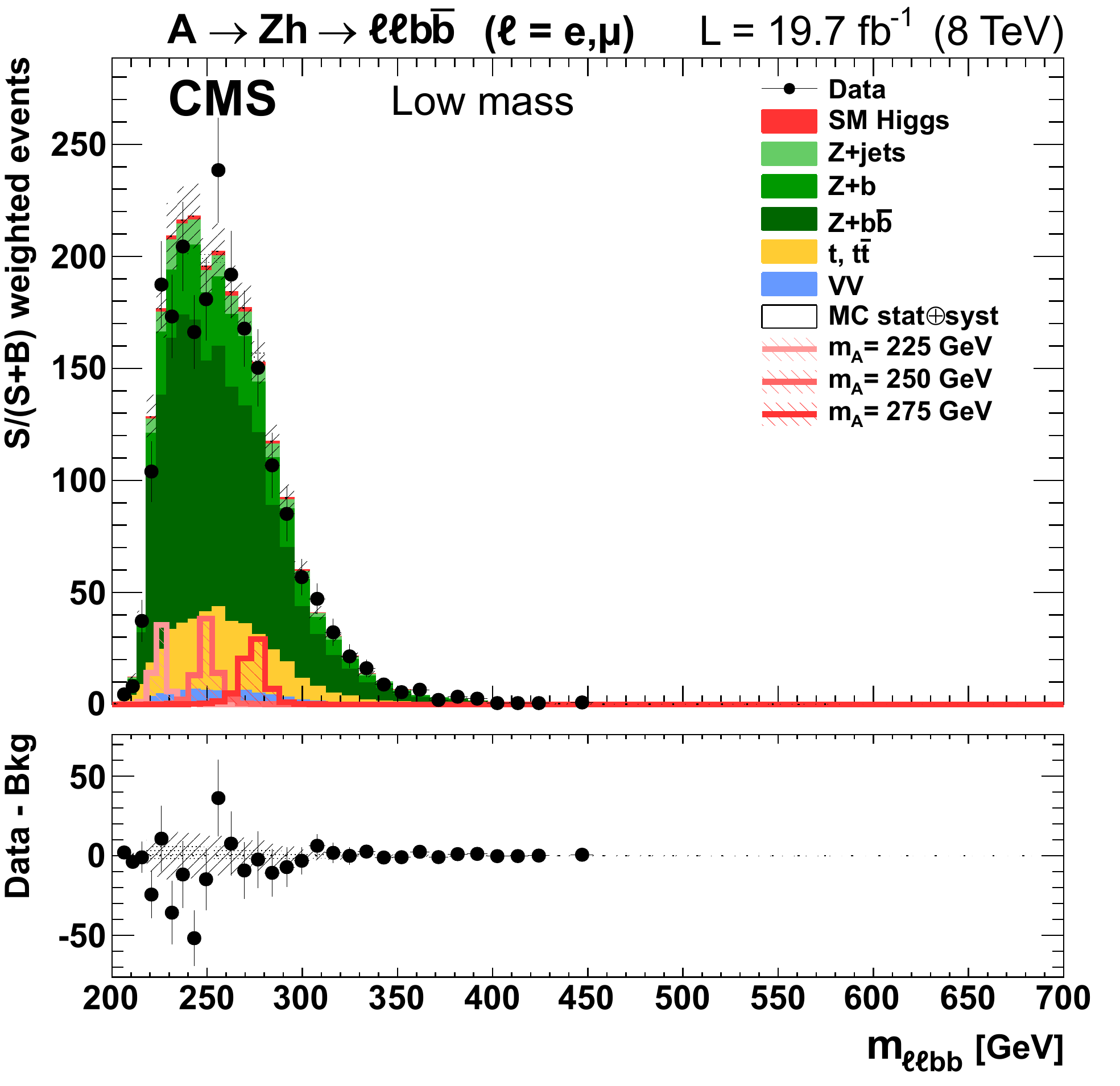}

  \includegraphics[width=0.95\cmsFigWidth]{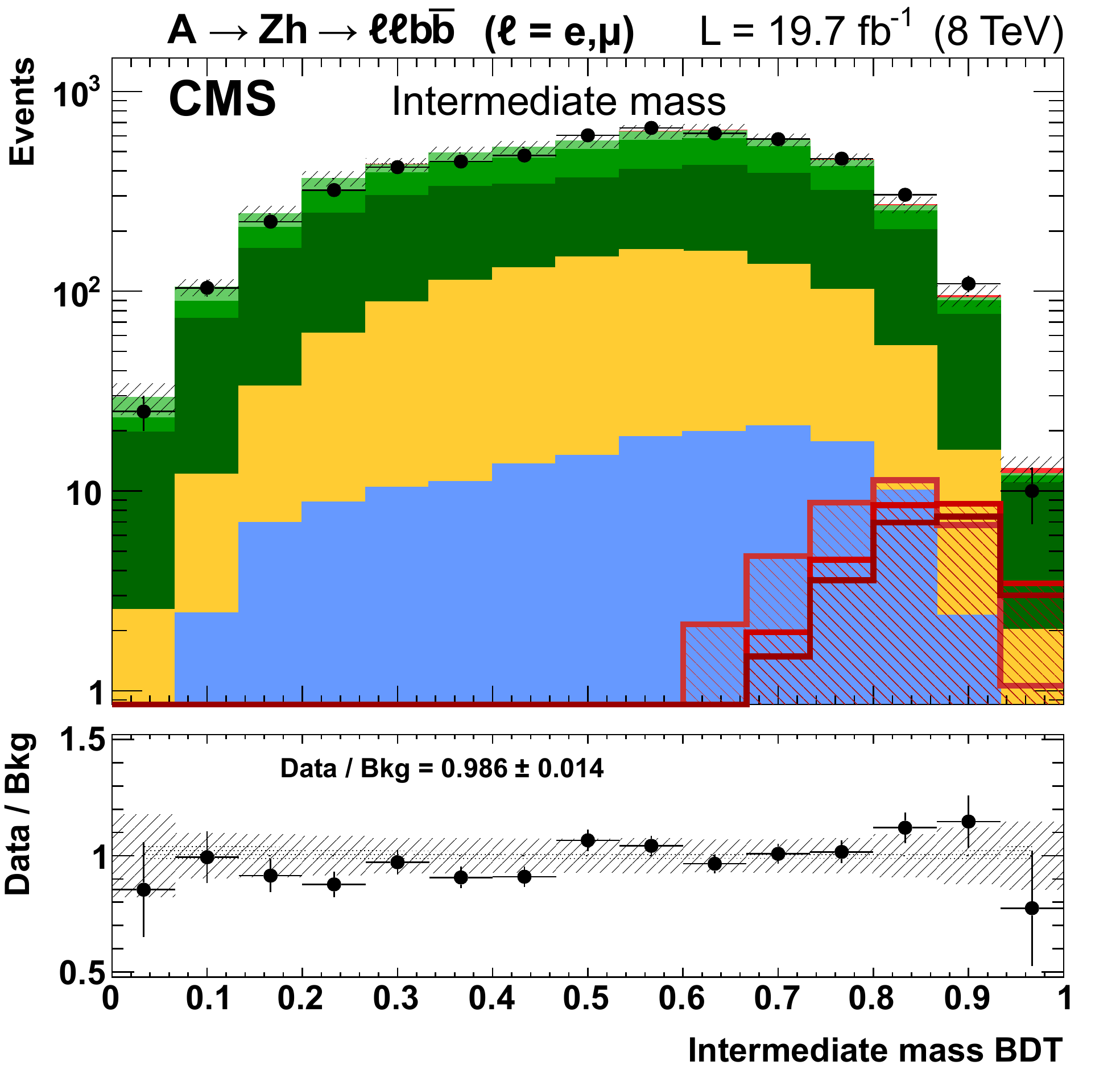}
  \includegraphics[width=0.95\cmsFigWidth]{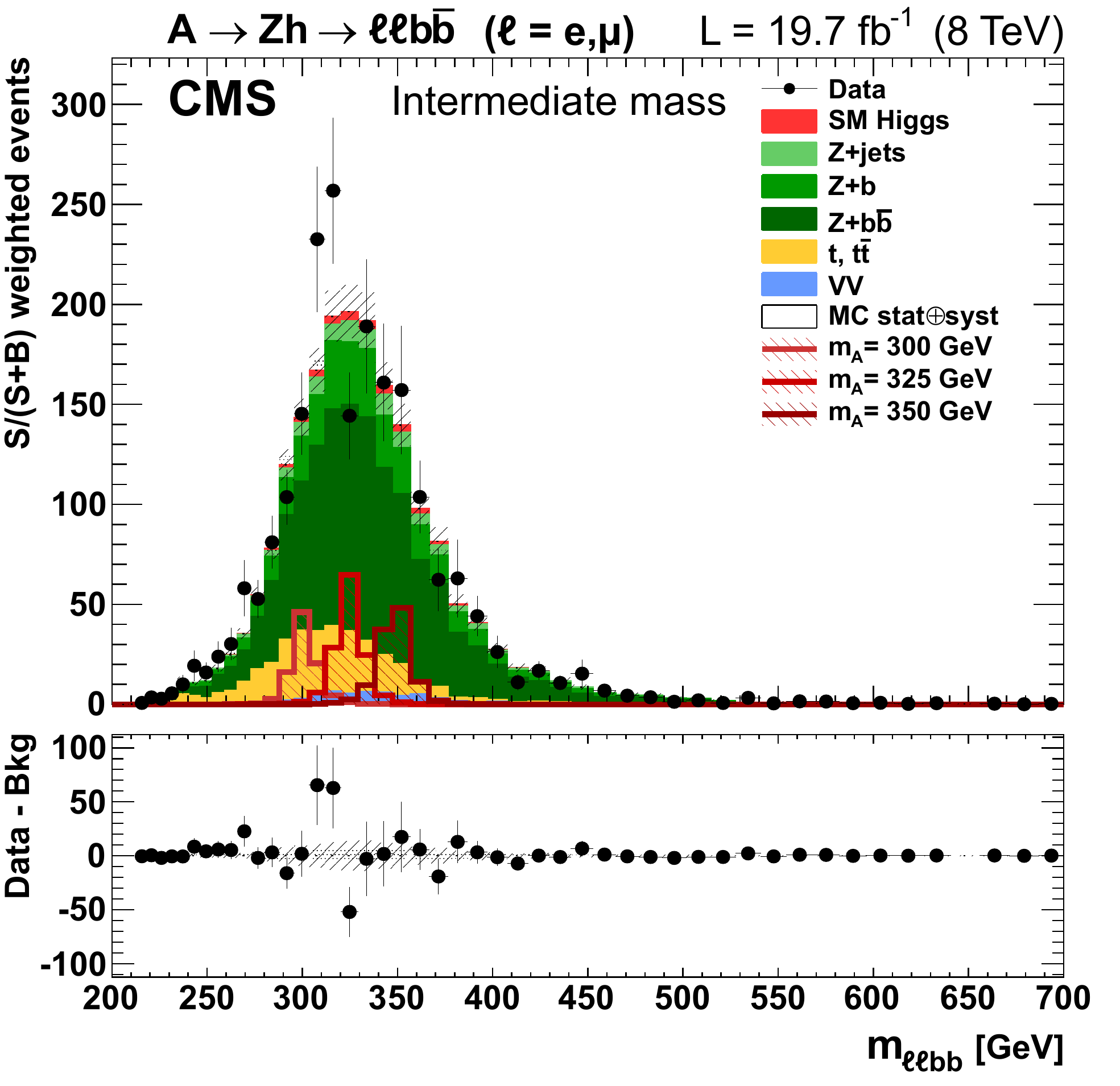}

  \includegraphics[width=0.95\cmsFigWidth]{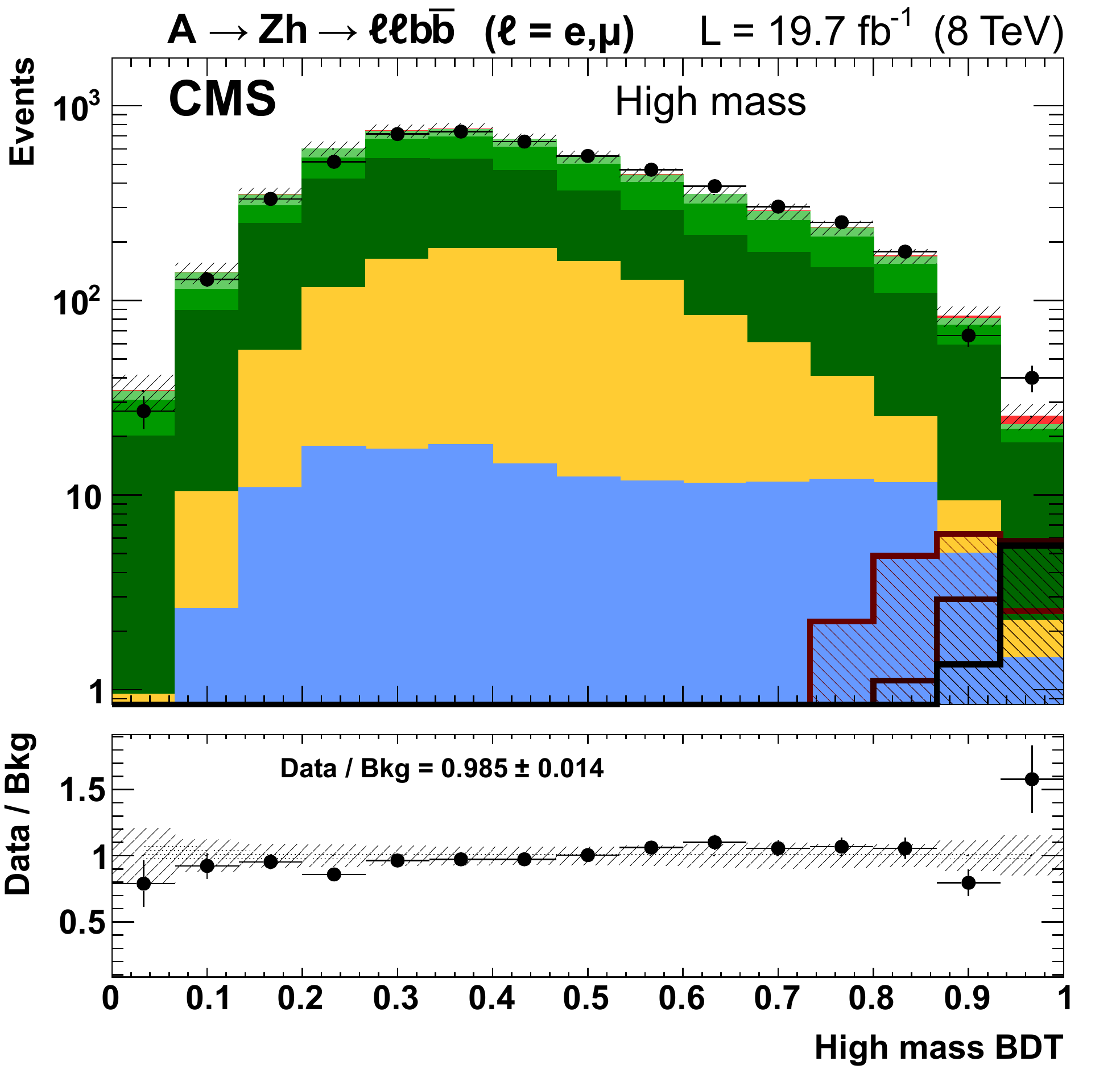}
  \includegraphics[width=0.95\cmsFigWidth]{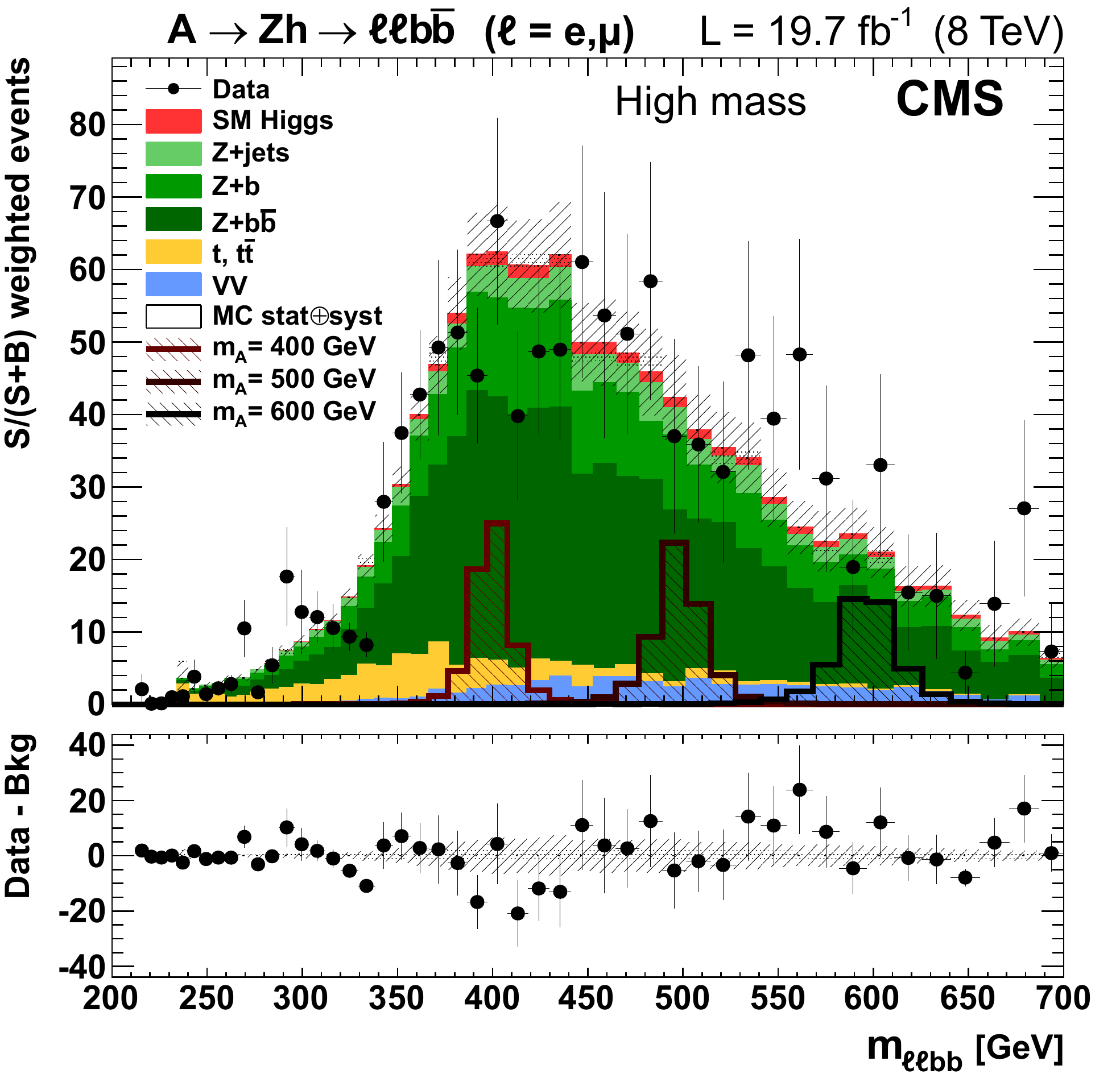}

  \caption{BDT outputs and invariant mass distributions in the low, intermediate, and high mass regions.
    The \mllbb plots are for $\mathrm{BDT}>0.6$, weighted by $\mathrm{S/(S+B)}$ in each BDT bin.
    Histograms for signal are normalized to the expected exclusion limit at 95\% confidence level.
    Statistical and systematic uncertainties in simulated samples are shown as well.
    Either the ratio (left) or the difference (right) between data and SM background is given at the bottom of each panel.
  }
  \label{fig:final}
\end{figure*}

Discrimination of signal from backgrounds is achieved through a
multivariate discriminant. Simulated mass points are divided into three
mass regions: low ($\mA=225$, 250, and 275\GeV), intermediate
($\mA=300$, 325, and 350\GeV), and high mass ($\mA=400$, 500, and 600\GeV).
Three boosted decision trees (BDT)~\cite{doi:97818609489850029} are
trained separately, one for each region.
The inputs of each BDT consist of 16 discriminating variables, selected from a list of more than 40 variables: the \pt of the \Z and \Ph boson candidates, the $\chi^2$ of the kinematic fit, the significance of \MET~\cite{1748-0221-10-02-P02006}, the dilepton invariant mass, the $\Delta{R}$ separation and the ``twist'' angle (defined as $\tan^{-1}{\Delta \varphi / \Delta \eta}$) between the two b jets~\cite{var}, their CSV discriminator values, the flight directions of the \Z boson and of the beam in the rest frame of the \A boson ($\cos \theta^*$),
the decay angle of the \Z boson relative to its flight direction in the rest frame of the \Z boson ($\cos\theta_1$), which is sensitive to the transverse polarization of the \Z boson along its flight direction, the angle of the pull vector~\cite{colors,PhysRevD.83.092002} of the highest-\pt jet, which exploits the color connection between the two b quarks originating from the \Ph boson,
the scalar sum of \MET and the \pt of jets and leptons in the event (\ST),
the number of jets with $\pt>20\GeV$, the event centrality, and aplanarity~\cite{var}.
The distribution in BDT outputs for data, for simulated signal (S), and for the expected SM background (B) events are shown in Fig.~\ref{fig:final}, weighting each entry in the \mllbb distribution by the expected S/(S$+$B) ratio of the BDT bin in the signal-sensitive region with BDT~$>0.6$.

\section{Systematic uncertainties}\label{sec:syst}

The sensitivity of the analysis is currently limited by the available data, and not by systematic uncertainties.

The uncertainties in the normalizations of the four main backgrounds (\Z{}+jets,
    \Zb, \Zbbbar, and \ttbar) originate both from the fits in the control regions and from the
extrapolation to the signal region. The former are reported in Table~\ref{tab:CR}, and the latter is
evaluated through a simultaneous likelihood fit to data and to simulated yields
in several statistically independent regions, obtained by altering the selections used to define the four
control regions.
A 13\% normalization uncertainty due to this extrapolation is assigned to \Z{}+jets, 12\%
to \Zb, 2.1\% to \Zbbbar, and 6.2\% to \ttbar events.
Normalization uncertainties for other SM backgrounds correspond to the ones on their measured
or theoretical cross sections.

The uncertainties in lepton reconstruction, identification, isolation, and
trigger efficiencies are evaluated through specific studies of events with masses in the region of the \Z peak.
Uncertainties in background or signal normalization and distribution due to uncertainties in jet
energy scale and resolution~\cite{CMS-PAS-JME-10-010} and b tagging scale factors~\cite{CMS-PAS-BTV-13-001} are estimated
by changing the corresponding values by ${\pm}1$ standard deviation ($\sigma$).
Additional systematic uncertainties affecting the normalization of backgrounds and signal
 from the choice of PDF~\cite{pdf,pdf2}, contributions from pileup, \MET
fluctuations because of the presence of unclustered energy in the event, and integrated
luminosity~\cite{lumi} are also considered in the analysis and are reported in Table~\ref{tab:Sys} (Normalization).

Results are extracted from an analysis based on a binned likelihood fit to the two-dimensional (2D) distribution of \mllbb versus BDT.
Dependence on jet energy scale and resolution, b tagging, as well as factorization and renormalization scales are propagated to the 2D-templates, taking into account the correlations between the two variables. The impact of reweighting the Monte Carlo distributions is considered as an additional source of background in modeling the uncertainty. Finally, the uncertainty from the limited number of simulated events is treated as in Ref.~\cite{Barlow1993219}. The sources of systematic uncertainty affecting the forms of the distributions are summarized in the first column of Table~\ref{tab:Sys} (Shape). The second and third columns indicate the respective ranges in the relative impact in percent, obtained by the changes implemented in the background and signal contributions.

The systematic uncertainty with the largest impact on the expected limit is from the reweighting of the background
 (accounting for a $\lesssim$6\% difference on the expected limit, depending on \mA), which is followed by the limitations in the number of simulated events ($\lesssim$4\%), and the factorization and renormalization scales ($\lesssim$2\%). The effect of the other sources is small ($\lesssim$1\%).

\begin{table*}[!htb]
  \centering
  \topcaption{Summary of systematic uncertainties. Normalization: sources of systematic uncertainty and their effect in percent on the normalization of signal and background distributions. Shape: sources of systematic uncertainty and the range of their effect in percent on relative changes made in the form of the background and signal distributions.}
  \label{tab:Sys}
  \begin{tabular}{lccc}
    \hline
    \multirow{2}{*}{Sources} & \multicolumn{2}{c}{Backgrounds} & Signal \\\cline{2-3}
              & Drell--Yan, \ttbar & Others & \\
    \hline
     \multicolumn{4}{c}{Normalization} \\
    \hline
    Control region fitting     & $<$2.4\%  & --- & ---            \\
    Extrapolation              & 2--13\%  & --- & ---            \\
    Lepton and trigger         & \multirow{2}{*}{---} & \multirow{2}{*}{2.5\%} & \multirow{2}{*}{2.5\%} \\
    \quad{}efficiency          &  & \\
    Jet energy scale           & --- & 5.7\% & 3.8--0.2\%  \\
    Jet energy resolution      & --- & 3.2\% & 0.8--0.5\%  \\
    b tagging                  & --- & 4.9\% & 3.6--3.2\%  \\
    Unclustered energy         & --- & 1.9\% & 1.4--1.0\%  \\
    Pileup                     & --- & 0.9\% & 1.2\%      \\
    PDF                        & --- & 4.3\% & 4.0--7.9\%  \\
    Cross section              & --- & 9.2--$15\%$ & ---       \\
    Integrated luminosity      & --- & 2.6\% & 2.6\%      \\
    \hline
    \multicolumn{4}{c}{Shape}\\
    \hline
    Jet energy scale           & \multicolumn{2}{c}{$<$4\%} & $<$8\% \\
    Jet energy resolution      & \multicolumn{2}{c}{$<$2\%} & $<$4\% \\
    b tagging                  & \multicolumn{2}{c}{$<$4\%} & $<$8\% \\
    Factorization and          & \multicolumn{2}{c}{\multirow{2}{*}{$<$6\%}} &\multirow{2}{*}{6--10\%} \\
    \quad{}renormalization scales     & \multicolumn{2}{c}{} \\
    Monte Carlo reweighting    & \multicolumn{2}{c}{$<$15\%} & --- \\
    Monte Carlo statistics     & \multicolumn{2}{c}{1--4\%} & --- \\
    \hline
  \end{tabular}
\end{table*}

\section{Results and their interpretation}\label{sec:results}

Results are obtained from the combined signal and background fit to the binned two-dimensional distribution of the four-body
invariant mass \mllbb and the BDT output in the signal-sensitive (BDT~$>$0.6) region.
With no evidence of significant deviation from background expectations,
the asymptotic modified frequentist method is used to determine the limit at the 95\% confidence level (CL) on the
contribution from signal, treating systematic uncertainties as nuisance parameters
that are integrated over in the fit~\cite{CLS1,CLS2,CMS-NOTE-2011-005,Asymptotic}.

\begin{figure}[!htb]\centering
    \includegraphics[width=\cmsFigWidthOneSix]{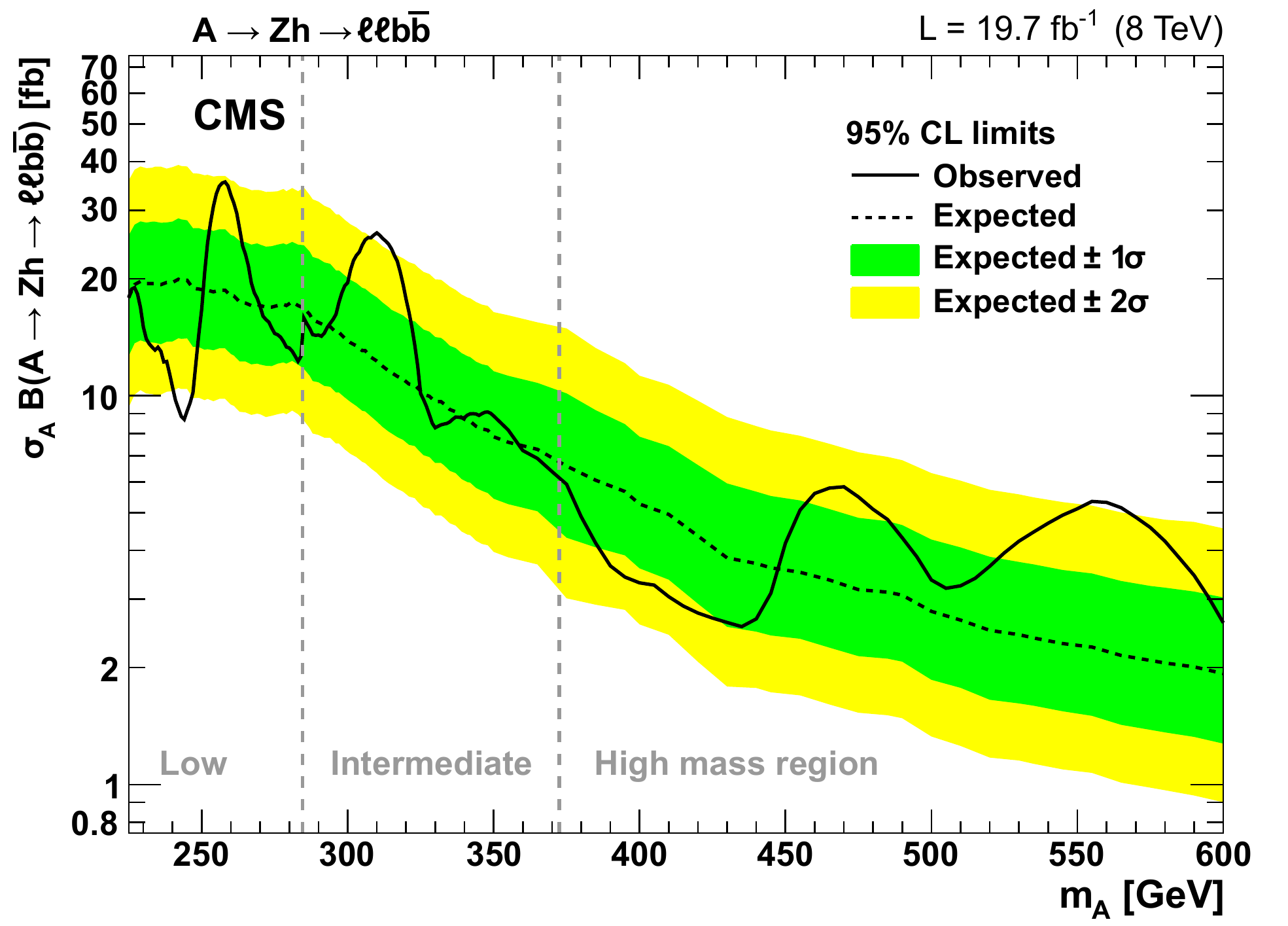}
  \caption{Observed and expected 95\% CL upper limit on
      $\sigma_\A \,\B(\Atollbb)$ as a function of \mA in the narrow-width
      approximation, including all statistical and systematic uncertainties.
      The green and yellow bands are the ${\pm}1$ and ${\pm}2\sigma$ uncertainty bands on the expected limit.
      }
  \label{fig:sigmaBRLimit}
\end{figure}

The observed limit, as well as the expected limit and its relative ${\pm}1$ and ${\pm}2 \sigma$ bands of uncertainty, are reported as a function of the \A boson mass in Fig.~\ref{fig:sigmaBRLimit} for $\sigma_\A \, \B(\Atollbb)$, \ie
the product of the cross section and the \AtoZh, \htobb, and \Ztoll branching fractions, with $\ell=\Pe$ or $\mu$.
The limits are obtained by considering the \A boson produced via the gluon-gluon fusion process in the narrow-width approximation.
Interpolated mass points are obtained as in Ref.~\cite{bib:th1fmorph}, and numerical values are reported in Table~\ref{tab:sysExp}.
A signal upper limit at 95\% CL is set on $\sigma_\A \,\B(\Atollbb)$, excluding
from 10 to 30\fb for \mA
near the kinematic threshold, $\approx$8\fb for $\mA \approx 2 \mt$,
and up to~$\approx$3\fb at the high end (600\GeV) of the considered mass range.
Comparable limits have been recently obtained for the same channel by the ATLAS Collaboration~\cite{Aad:2015wra}.
The most significant excess at $\mllbb=560\GeV$ has a local significance of
$2.6 \sigma$, which becomes a global $1.1 \sigma$ significance taking
into account the \emph{look-elsewhere} effect~\cite{grossvitells}.

\begin{figure}[!htb]\centering
    \includegraphics[width=\cmsFigWidth]{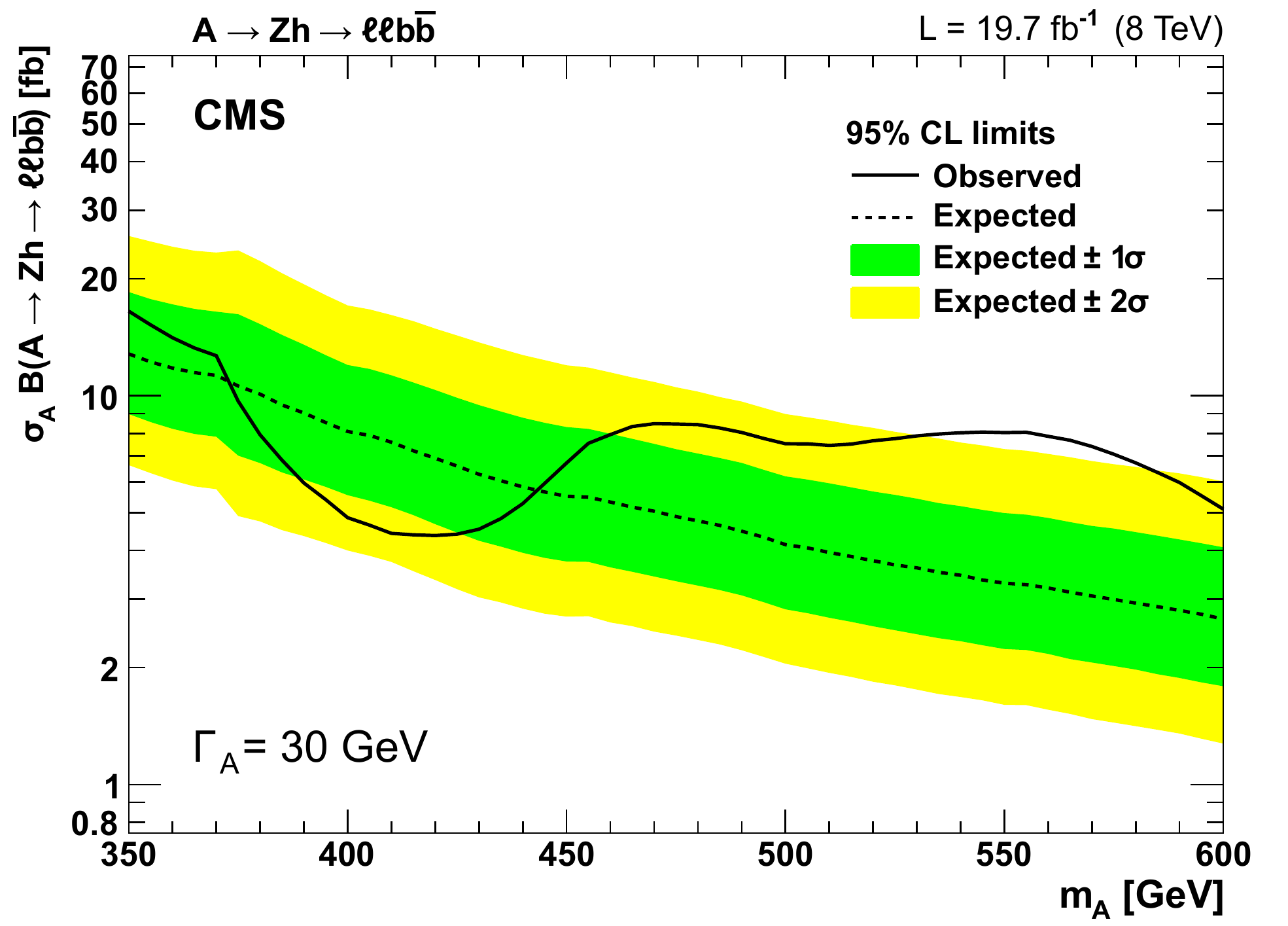}
    \includegraphics[width=\cmsFigWidth]{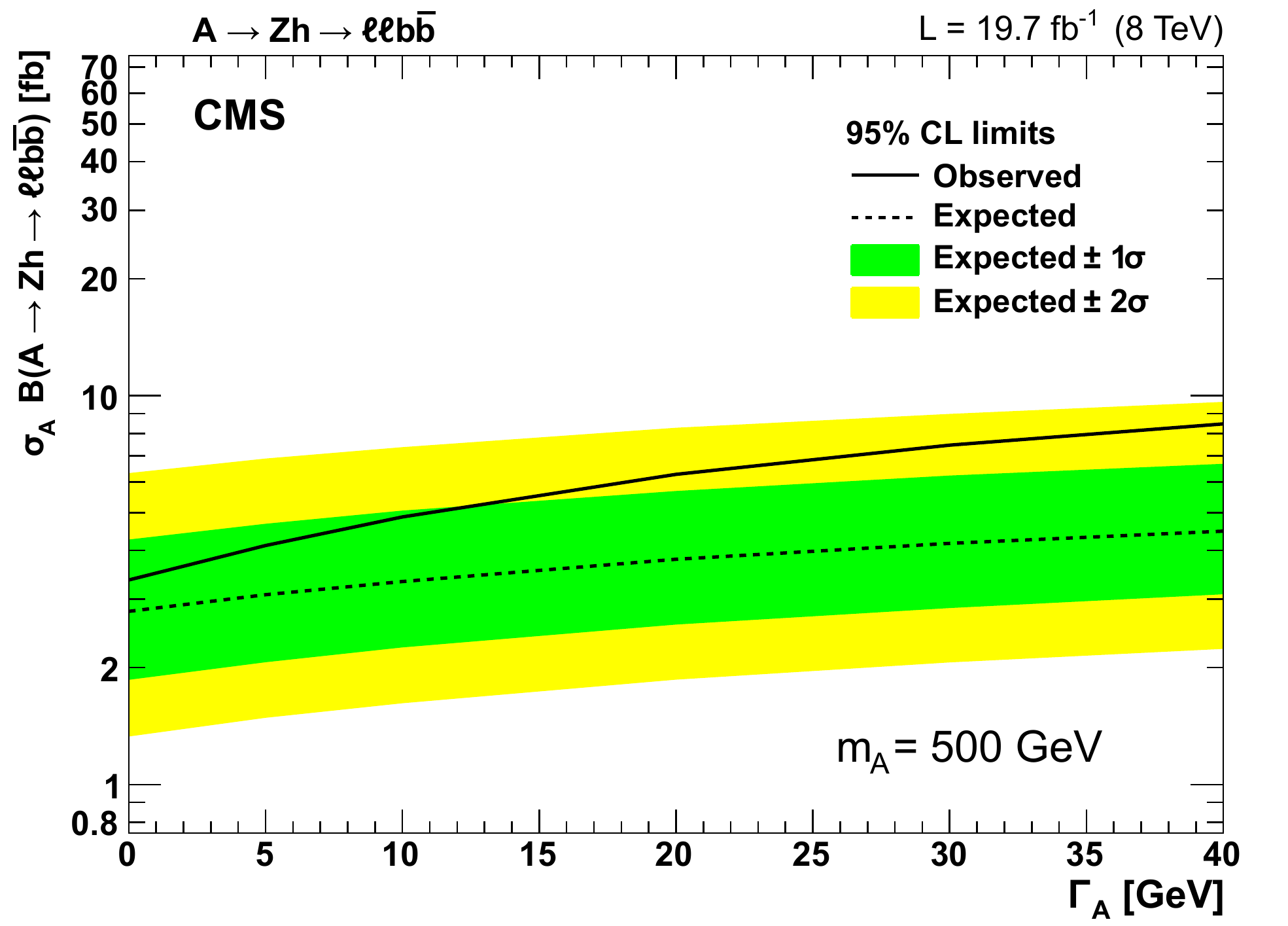}
  \caption{Observed and expected 95\% CL upper limit on
      $\sigma_\A \,\B(\Atollbb)$ for $\Gamma_\A=30\GeV$ as a function of \mA
      (\cmsLeft), and for $\mA=500\GeV$ as a function of the width of the \A
      boson (\cmsRight).}
  \label{fig:sigmaBRWidth}
\end{figure}

For $\mA>2 \mt$, the width of the \A boson depends strongly on the model parameters.
Different limits are provided by taking into account the natural width of the \A boson ($\Gamma_\A$)
in the reconstructed \mllbb, leaving the BDT unchanged.
Figure~\ref{fig:sigmaBRWidth} shows the exclusion limit above $2 \mt$
for an average width of 30\GeV, and the dependence
of the observed limit on $\Gamma_\A$ for $\mA= 500\GeV$.
The local and global significance at $\mllbb= 560\GeV$ become, respectively, $2.9\sigma$ and $1.5\sigma$,
assuming $\Gamma_\A=30\GeV$.

As an independent cross-check of the 2D fit, the signal is extracted by
applying two complementary strategies, based on one-dimensional
fits. The first one consists of fitting the \mllbb distribution, after
selecting events in a signal-enriched region by applying a $\mathrm{BDT}>0.8$
selection. The second relies on fits to the BDT distributions, after
selecting events within the resolution of the signal \mllbb peak.
The two methods give upper limits compatible with those from the 2D fit, but 10 to 20\% less stringent.

The results are interpreted in terms of Type-I and Type-II
2HDM formulations~\cite{bib:Branco20121}.
The $\B(\AtoZh)$ and $\B(\htobb)$ branching fractions
and the signal cross sections are computed at
next-to-next-to-leading-order (NNLO) with \textsc{SuShi
    1.2.0}~\cite{Harlander20131605} and \textsc{2hdmc
    1.6.4}~\cite{Eriksson2010189}, respectively, using the
    MSTW2008LO, NLO, and NNLO sets of PDF.
The $\B(\Ztoll)$ branching fraction, with $\ell=\Pe$ or $\mu$, is taken from the measured value~\cite{Agashe:2014kda}.
Both gluon-gluon fusion and associated production with b quarks have been considered.
The latter is rescaled to the fusion process, taking account of the difference
in acceptance for signal, as well as the efficiency for selecting dijet pairs
in the presence of combinatorial contributions from additional b quarks in the
event.
The parameters used for the models are: $\mh=125\GeV$, $\mH=\mHpm=\mA$,
$m^2_{12}=\mA^2 [\tan\beta]/[1+\tan^2\beta]$, $\lambda_{6,7}=0$~\cite{Harlander:2013qxa}, while
$\mA=225$--600\GeV, $0.1\le\tanb\le100$, and $-1\le\cosba\le1$, using the
convention $0<\beta-\alpha<\pi$, where \tanb and $\alpha$ are,
respectively, the ratio of the vacuum expectation values, and the mixing angle
of the two Higgs doublets~\cite{bib:Branco20121}.

\begin{figure}[thp]\centering
    \includegraphics[width=\cmsFigWidth]{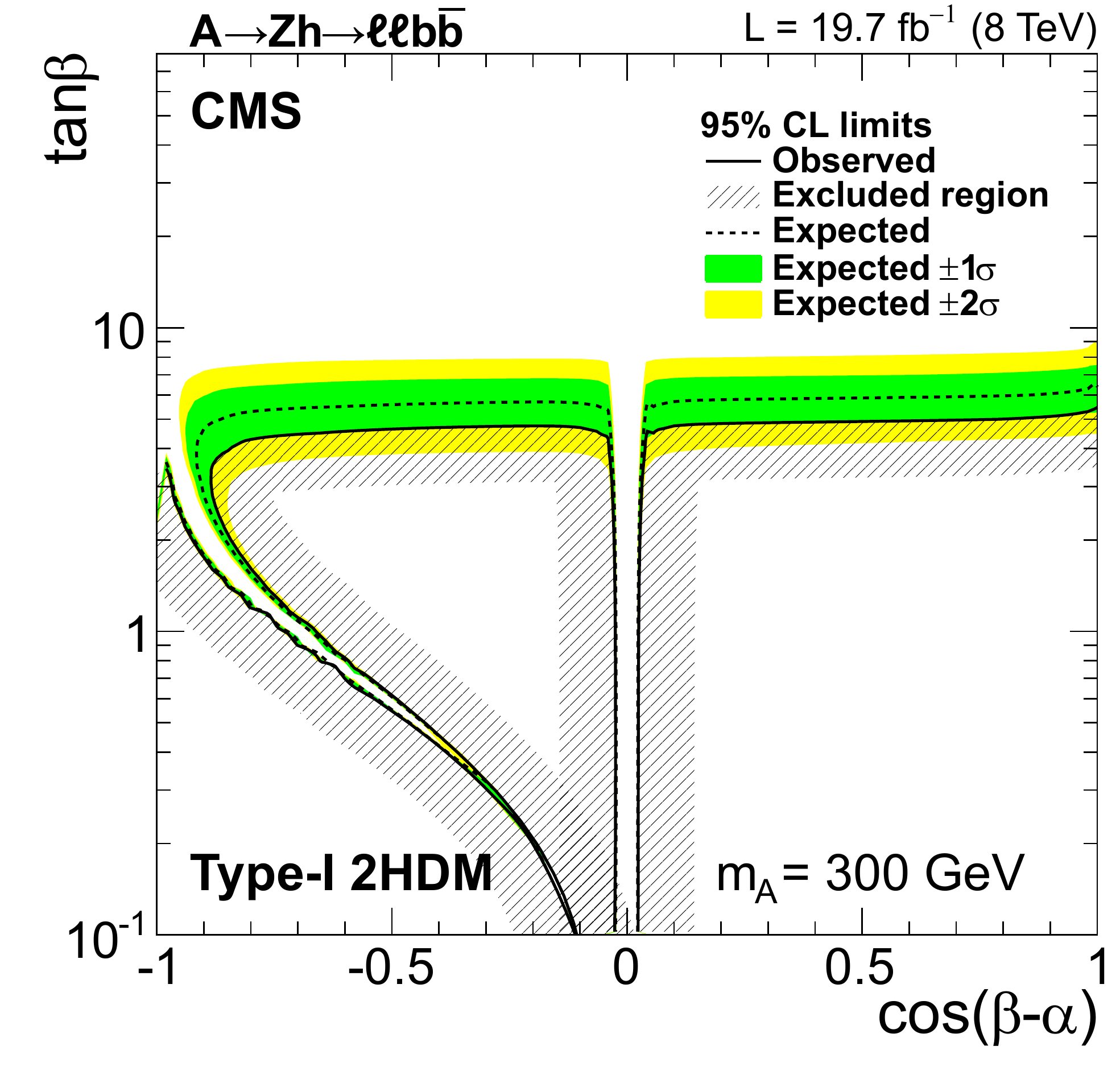}
    \includegraphics[width=\cmsFigWidth]{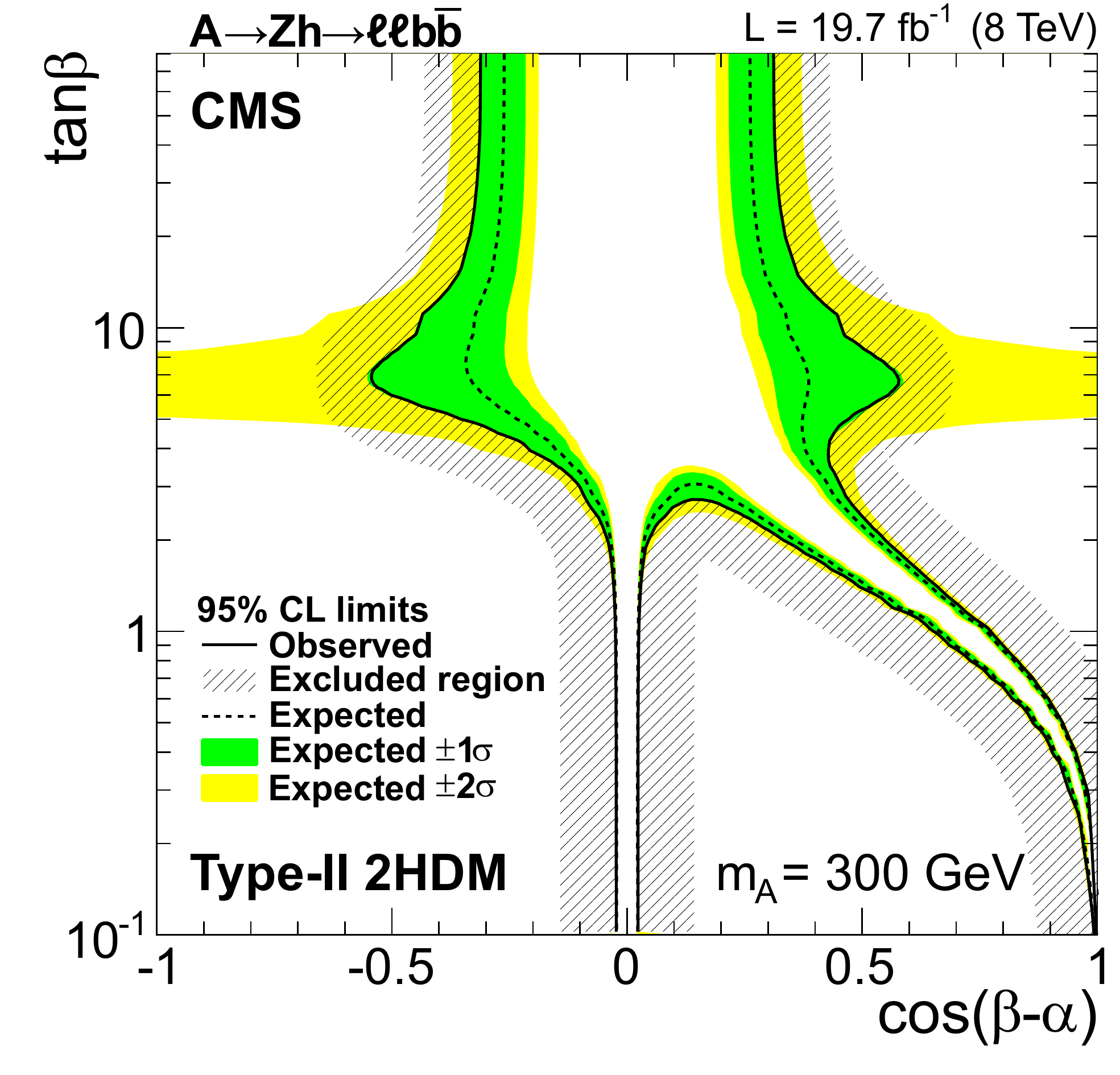}

   \caption{Observed and expected (together with ${\pm}1,\,2\sigma$ uncertainty bands) exclusion limit for Type-I (\cmsLeft) and Type-II (\cmsRight) models, as a function of \tanb and \cosba.
   Contours are derived from the projection on the 2HDM parameter space for the \mA = 300\GeV signal hypothesis; the observed limit is close to $1 \sigma$ above the expected limit, as shown in Fig.~\ref{fig:sigmaBRLimit}.
  }
  \label{fig:2HDM}
\end{figure}

The observed limit, together with the expected limit and
its relative ${\pm}1$ and ${\pm}2 \sigma$ uncertainty bands,
are shown in Fig.~\ref{fig:2HDM}, interpreted in Type-I and
Type-II 2HDM for \mA=\xspace300\GeV. A sizeable fraction of the 2HDM phase space is excluded at a 95\% CL with respect to the previous CMS searches~\cite{bib:Khachatryan:2014jya}.

\begin{table*}[!htb]\centering
  \topcaption{Observed and expected 95\% CL upper limits on $\sigma_\A \,\B(\Atollbb)$ as a function of \mA in the narrow-width approximation, including statistical and systematic uncertainties.}
  \begin{tabular}{lccc ccc ccc}

    \hline
    $\mA\,[\GeVns{}]$    & 225  &  250 &  275 &  300 &  325 &  350 &  400 &  500 & 600 \\
    \hline
    Observed [fb]  & 17.9 & 16.8 & 14.8 & 19.5 & 10.1 & 8.84 & 3.29 & 3.35 & 2.61 \\

    Expected [fb]  & 17.9 & 18.1 & 16.4 & 13.6 & 10.0 & 7.84 & 5.27 & 2.79 & 1.93 \\
    \hline

  \end{tabular}
  \label{tab:sysExp}
\end{table*}

\section{Summary}\label{sec:conclusion}

A search is presented for new physics in the extended Higgs sector, in signatures expected from decays of a pseudoscalar boson \A into a \Z boson and an SM-like \Ph boson, with the \Z boson decaying into $\ell^+\ell^-$ ($\ell$ being either \Pe~or $\mu$) and the \Ph boson into \bbbar. Different techniques are employed to increase the sensitivity to signal, exploiting the presence of the three resonances \A, \Z, and \Ph to discriminate against standard model backgrounds.
Upper limits at a 95\% CL are set on the product of a narrow pseudoscalar boson cross section and branching fraction $\sigma_\A \,\B(\Atollbb)$, which exclude 30 to 3\fb at the low and high ends of the 250--600\GeV mass range.
Results are also presented as a function of the width of the \A boson.
Interpretations are given in the context of Type-I and Type-II 2HDM formulations, thereby reducing the parameter space for extensions of the standard model.

\begin{acknowledgments}
We congratulate our colleagues in the CERN accelerator departments for the excellent performance of the LHC and thank the technical and administrative staffs at CERN and at other CMS institutes for their contributions to the success of the CMS effort. In addition, we gratefully acknowledge the computing centers and personnel of the Worldwide LHC Computing Grid for delivering so effectively the computing infrastructure essential to our analyses. Finally, we acknowledge the enduring support for the construction and operation of the LHC and the CMS detector provided by the following funding agencies: BMWFW and FWF (Austria); FNRS and FWO (Belgium); CNPq, CAPES, FAPERJ, and FAPESP (Brazil); MES (Bulgaria); CERN; CAS, MoST, and NSFC (China); COLCIENCIAS (Colombia); MSES and CSF (Croatia); RPF (Cyprus); MoER, ERC IUT and ERDF (Estonia); Academy of Finland, MEC, and HIP (Finland); CEA and CNRS/IN2P3 (France); BMBF, DFG, and HGF (Germany); GSRT (Greece); OTKA and NIH (Hungary); DAE and DST (India); IPM (Iran); SFI (Ireland); INFN (Italy); MSIP and NRF (Republic of Korea); LAS (Lithuania); MOE and UM (Malaysia); CINVESTAV, CONACYT, SEP, and UASLP-FAI (Mexico); MBIE (New Zealand); PAEC (Pakistan); MSHE and NSC (Poland); FCT (Portugal); JINR (Dubna); MON, RosAtom, RAS and RFBR (Russia); MESTD (Serbia); SEIDI and CPAN (Spain); Swiss Funding Agencies (Switzerland); MST (Taipei); ThEPCenter, IPST, STAR and NSTDA (Thailand); TUBITAK and TAEK (Turkey); NASU and SFFR (Ukraine); STFC (United Kingdom); DOE and NSF (USA).

Individuals have received support from the Marie-Curie program and the European Research Council and EPLANET (European Union); the Leventis Foundation; the A. P. Sloan Foundation; the Alexander von Humboldt Foundation; the Belgian Federal Science Policy Office; the Fonds pour la Formation \`a la Recherche dans l'Industrie et dans l'Agriculture (FRIA-Belgium); the Agentschap voor Innovatie door Wetenschap en Technologie (IWT-Belgium); the Ministry of Education, Youth and Sports (MEYS) of the Czech Republic; the Council of Science and Industrial Research, India; the HOMING PLUS program of the Foundation for Polish Science, cofinanced from European Union, Regional Development Fund; the Compagnia di San Paolo (Torino); the Consorzio per la Fisica (Trieste); MIUR project 20108T4XTM (Italy); the Thalis and Aristeia programs cofinanced by EU-ESF and the Greek NSRF; and the National Priorities Research Program by Qatar National Research Fund.
\end{acknowledgments}

\bibliography{auto_generated}

\cleardoublepage \appendix\section{The CMS Collaboration \label{app:collab}}\begin{sloppypar}\hyphenpenalty=5000\widowpenalty=500\clubpenalty=5000\textbf{Yerevan Physics Institute,  Yerevan,  Armenia}\\*[0pt]
V.~Khachatryan, A.M.~Sirunyan, A.~Tumasyan
\vskip\cmsinstskip
\textbf{Institut f\"{u}r Hochenergiephysik der OeAW,  Wien,  Austria}\\*[0pt]
W.~Adam, E.~Asilar, T.~Bergauer, J.~Brandstetter, E.~Brondolin, M.~Dragicevic, J.~Er\"{o}, M.~Flechl, M.~Friedl, R.~Fr\"{u}hwirth\cmsAuthorMark{1}, V.M.~Ghete, C.~Hartl, N.~H\"{o}rmann, J.~Hrubec, M.~Jeitler\cmsAuthorMark{1}, V.~Kn\"{u}nz, A.~K\"{o}nig, M.~Krammer\cmsAuthorMark{1}, I.~Kr\"{a}tschmer, D.~Liko, T.~Matsushita, I.~Mikulec, D.~Rabady\cmsAuthorMark{2}, B.~Rahbaran, H.~Rohringer, J.~Schieck\cmsAuthorMark{1}, R.~Sch\"{o}fbeck, J.~Strauss, W.~Treberer-Treberspurg, W.~Waltenberger, C.-E.~Wulz\cmsAuthorMark{1}
% ...
\vskip\cmsinstskip
% ...
1:~~Also at Vienna University of Technology, Vienna, Austria\\
2:~~Also at CERN, European Organization for Nuclear Research, Geneva, Switzerland\\
% ...

\end{sloppypar}
\end{document}